\documentclass[twocolumn,
superscriptaddress,amssymb,amsmath,amsfonts,
 aps,
pra,
]{revtex4-2}

\usepackage{graphicx}
\usepackage{dcolumn}
\usepackage{bm}
\usepackage{hyperref}
\usepackage{braket}
\usepackage{xcolor}
\usepackage[export]{adjustbox}
\DeclareMathOperator{\Tr}{Tr}

\begin{document}

\title{Delocalization of quantum information in long-range interacting systems}
\author{Darvin Wanisch}
\email{darvin.wanisch@uni-jena.de}
\affiliation{Theoretisch-Physikalisches Institut, Friedrich-Schiller-Universität Jena, Max-Wien-Platz 1, 07743 Jena, Germany}
\affiliation{Helmholtz-Institut Jena, Fr\"obelstieg 3, 07743 Jena, Germany}

\author{Stephan Fritzsche}
\affiliation{Theoretisch-Physikalisches Institut, Friedrich-Schiller-Universität Jena, Max-Wien-Platz 1, 07743 Jena, Germany}
\affiliation{Helmholtz-Institut Jena, Fr\"obelstieg 3, 07743 Jena, Germany}
\affiliation{GSI Helmholtzzentrum für Schwerionenforschung GmbH, Planckstra\ss e 1, 64291 Darmstadt, Germany}
\date{\today}

\begin{abstract}
We investigate the delocalization of quantum information in the nonequilibrium dynamics of the $XY$ spin chain with asymptotically decaying interactions $\sim 1/r^{\alpha}$. As a figure of merit, we employ the tripartite mutual information (TMI), whose sign indicates if quantum information is predominantly shared globally. Interestingly, the sign of the TMI distinguishes regimes of the exponent $\alpha$ that are known for different behaviour of information propagation. While an effective causal region bounds the propagation of information, if interactions decay sufficiently fast, this information is mainly delocalized, which leads to the necessity of global measurements. Furthermore, the results indicate that mutual information is monogamous for all possible partitionings in this case, implying that quantum entanglement is the dominant correlation. If interactions decay sufficiently slow, though information can propagate (quasi-)instantaneously, it is mainly accessible by local measurements at early times.  Furthermore, it takes some finite time until correlations start to become monogamous, which suggests that entanglement is not the dominant correlation at early times. Our findings give new insights into the dynamics, and structure of quantum information in many-body systems with long-range interactions, and might get verified on state-of-the-art experimental platforms.
\end{abstract}

\maketitle

\section{Introduction}

Entanglement, a key resource in quantum information processing \cite{nielsen}, is believed to give precious insights into exciting physical phenomena in a variety of fields \cite{rev_dynqi_lewisswan}. Driven by recent progress on experimental platforms such as trapped ions \cite{sim_ion}, ultracold atoms in optical lattices \cite{sim_gases,sim_gases2}, and Rydberg atoms \cite{sim_rydberg,sim_rydberg2}, quantum many-body systems out of equilibrium gained a lot of attention lately \cite{rev_neq_eisert}. How quantum information propagates, and distributes over the degrees of freedom of a many-body system is fundamental for a plethora of subjects, ranging from the simulability of these systems on classical computers \cite{simulability_schuch}, to the AdS/CFT correspondance \cite{gravity_qi}. 

A few decades ago, Lieb and Robinson \cite{lieb_robinson} proved the emergence of a causal region, or lightcone, in nonrelativistic systems with sufficiently local interactions. That is to say: there exists a finite speed $v_\mathrm{LR}$ at which correlations and, therefore, information can propagate through the system. In consequence, correlations among remote regions get exponentially suppressed until some time that is proportional to their distance, $t\sim r$. Today's experimental platforms offer vast possibilities to explore the physics of quantum lattice models with increasing accuracy and system size. In some cases, the systems' constituents couple through long-range forces, which can result in interactions proportional to a powerlaw $\sim 1/r^{\alpha}$. If the exponent $\alpha$ is small, interactions are not sufficiently local anymore. As these platforms aim to serve as reliable quantum simulators \cite{feynman_sim} in the future, this led to a renewed theoretical interest in many-body systems with long-range interactions.

It is particularly intriguing that the notion of causality does not necessarily apply to long-range interacting systems, and information propagation can differ substantially \cite{lr_spread_hauke,lr_spread_schachenmayer,eisert_break_locality}. While the causal region continuously alters with decreasing $\alpha$, it even can be absent if the exponent of the powerlaw decay is smaller than the systems' dimension, $\alpha<D$. Thus, information can propagate (quasi-)instantaneously between remote regions. Counterintuitively, this regime was further associated with slower growth of bipartite entanglement \cite{lr_spread_schachenmayer,lr_entgrowth_lerose}. The exponent of the powerlaw decay, therefore, strongly affects the nonequilibrium properties of a many-body system, and subsequent works aimed to improve the understanding of information propagation in long-range interacting systems \cite{lightcones_lr_fossfeig,lr_spread_buyskikh,lr_area_law_gong,lieb_robinson_imp_else}. Here, we want to shed more light on the particular structure of quantum information. Especially, we investigate how the exponent $\alpha$ determines if quantum information is \textit{delocalized}, i.e.,  if information is rather shared globally than among individual subsystems.

This work is structured as follows. In Sec.\,\ref{sec:entropy_corr}, we introduce the quantities we use to probe the correlation structure of a many-body system, namely the Von Neumann entropy, the mutual information, and the tripartite mutual information (TMI). The TMI accounts for the delocalization of quantum information. Thereafter, in Sec.\,\ref{sec:deloc}, we investigate the TMI in the nonequilibrium dynamics of the long-range $XY$ spin chain. At first, we provide analytical arguments that quantum information does not delocalize in the 1-excitation subspace, which can be understood by the limited amount of entanglement that emerges. Subsequently, we numerically investigate the dynamics in the largest subspace of the model. In this case, we deduce regimes for the exponent $\alpha$ which exhibit different structure of quantum information at early times. How one can test our findings in the laboratory is briefly touched in Sec. \ref{sec:exp}. Finally, in Sec. \ref{sec:diss}, we discuss our results, pose some open questions, and talk about future directions. 

\section{\label{sec:entropy_corr}Entropy and correlation measures}

Let us establish what quantities we henceforth use to probe the correlation structure of a quantum many-body system. To capture correlations as a property of the quantum state itself, we are mainly concerned with operator-independent quantities. We shall, therefore, introduce the fundamental quantity in this regard: the \textit{Von Neumann entropy}. Given a quantum system $\Omega$ that is described by a pure state $\rho=\Ket{\Psi}\Bra{\Psi}$, associated with some finite-dimensional Hilbert space $\mathfrak{H}$, the Von Neumann entropy for a subsystem $A\subset\Omega$ is defined as
\begin{align}
    \label{entropy}
    S_A=-\Tr\left[\rho_A\log\left(\rho_A\right)\right]=-\sum_{\lambda\in\sigma\left(\rho_A\right)}\lambda\log\left(\lambda\right)\,\text{.}
\end{align}
In Eq.\,\eqref{entropy}, $\rho_A=\Tr_{\bar{A}}\left(\Ket{\Psi}\Bra{\Psi}\right)$ is the reduced state associated with $A$, $\sigma\left(\rho_A\right)$ denotes the spectrum of $\rho_A$, and $\bar{A}=\Omega\setminus A$ is the complement of $A$. The logarithm is taken to base 2, if not mentioned otherwise.

\begin{figure}
    \includegraphics[width=.15\textwidth,valign=t]{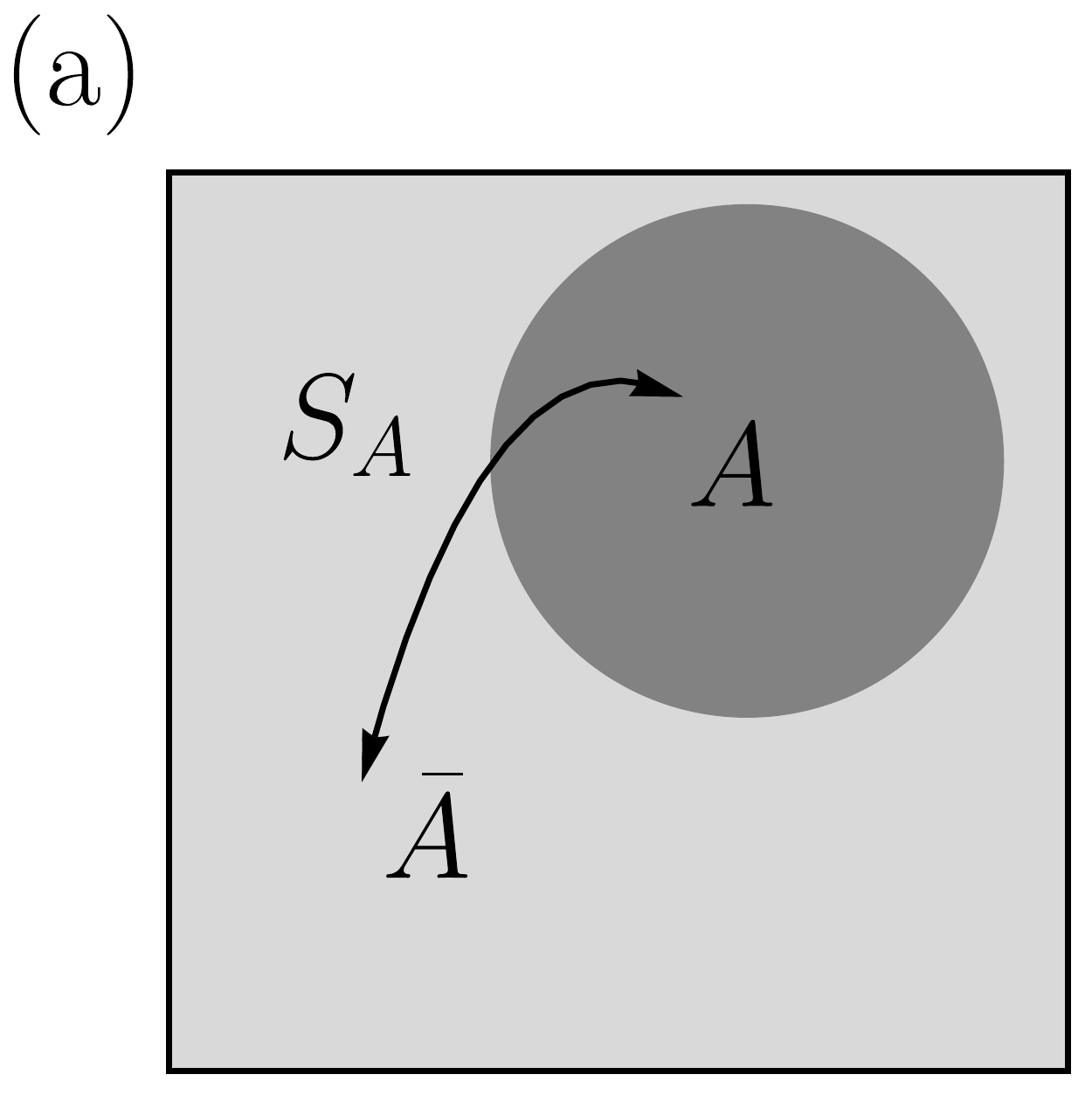}
    \hspace{.005\textwidth}
    \includegraphics[width=.15\textwidth,valign=t]{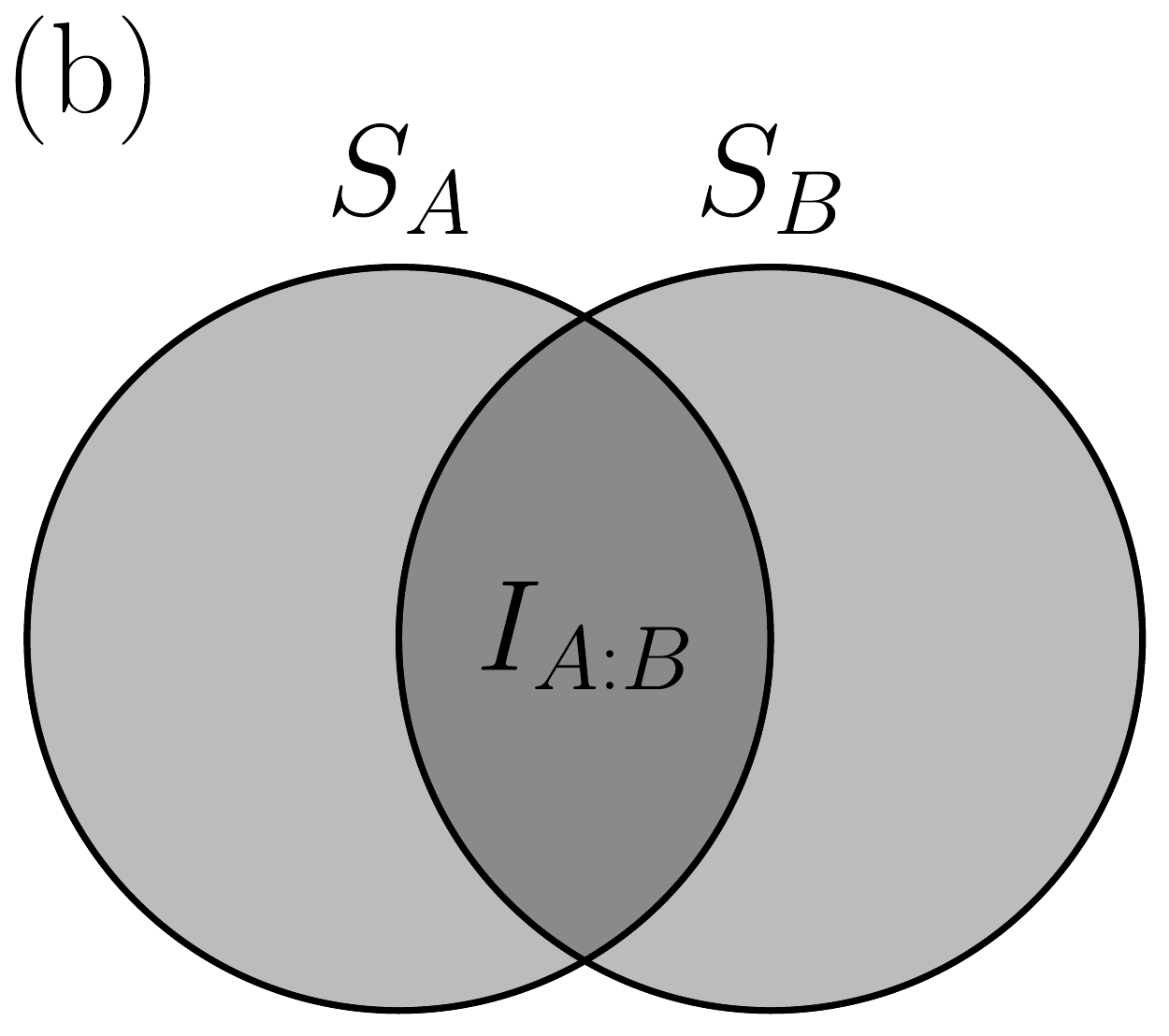}
    \hspace{.005\textwidth}
    \includegraphics[width=.15\textwidth,valign=t]{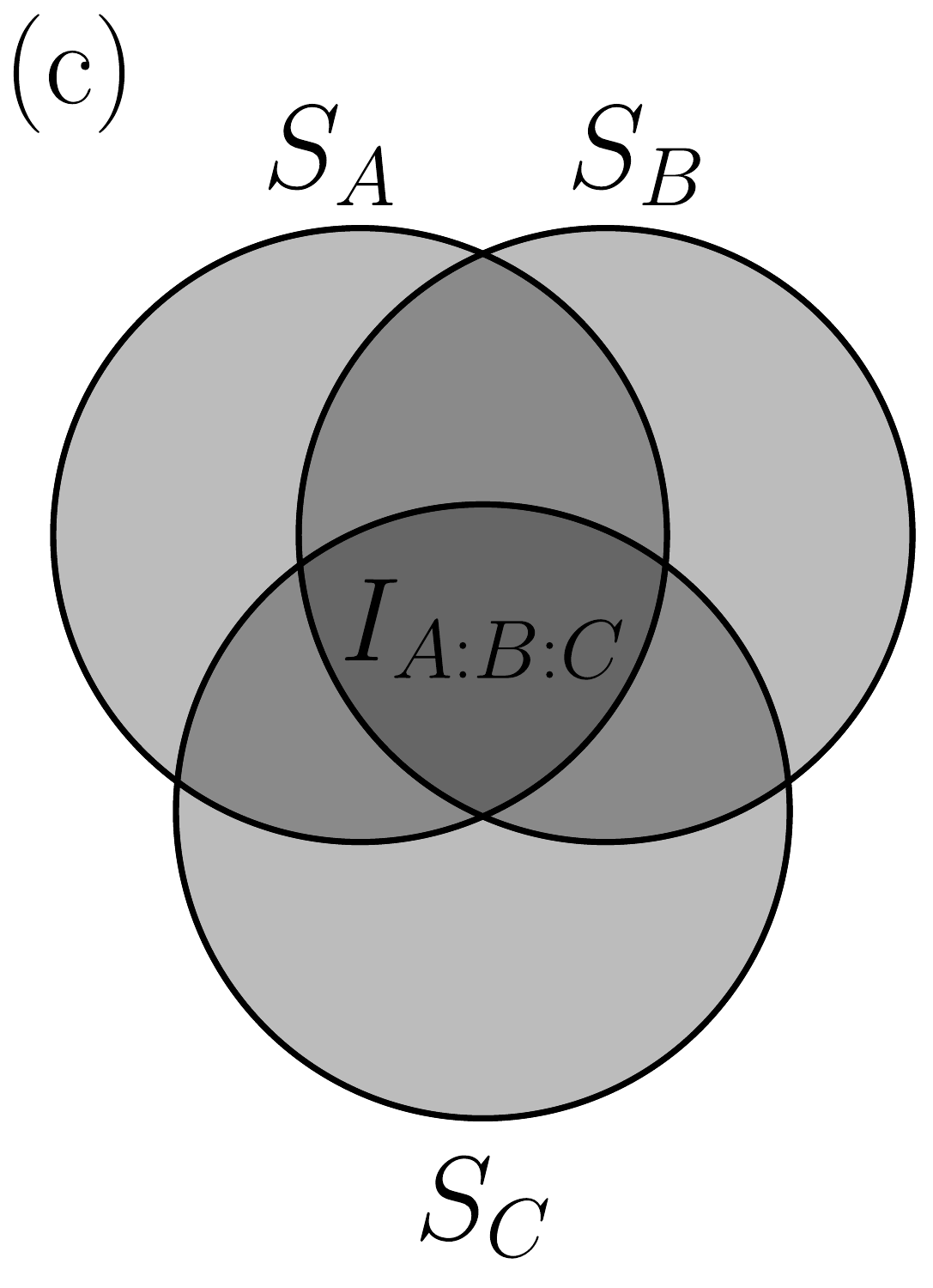}
    \caption{Correlations in a closed quantum system $\Omega\,$: (a) The Von Neumann entropy \eqref{entropy} of some subsystem $A\subset\Omega$ measures the amount of entanglement between this subsystem and its environment. (b) Given two subsystems $A$ and $B$, the mutual information \eqref{mi} determines their total correlation, which is illustrated as the intersection of both entropies in the Venn diagram. (c) The TMI \eqref{tmi} probes the distribution of quantum information among three subsystems $A,B$, and $C$. In the language of Venn diagrams, this corresponds to the intersection of all three entropies. A large negative value suggests that most information about $A$ is only accessible by joint measurements on $BC$.}
    \label{fig:corr_measures}
\end{figure}

Since the overall system $\Omega$ has zero entropy, a nonvanishing entropy for subsystem $A$ indicates that we lose information about the state of $A$ if we neglect its environment $\bar{A}$. Hence, Eq.\,\eqref{entropy} is a natural measure of entanglement between subsystem $A$ and its environment $\bar{A}$, see Fig\,\ref{fig:corr_measures}(a). The entropy is in general bounded by
\begin{align}
    \label{vne_bound_gen}
    0\leq S_A \leq \log\left[\min\left(d_A,d_{\bar{A}}\right)\right]\,\text{,}
\end{align}
where $d_{A}$ is the dimension of the Hilbert space associated with $A$. The right equality in \eqref{vne_bound_gen} holds iff $A$ is maximally entangled with $\bar{A}$, while the left equality  holds iff the systems’ state is separable with respect to the bipartition $\left\{A,\bar{A}\right\}$, i.e.,  $\Ket{\Psi}=\Ket{\Psi}_A\otimes\Ket{\Psi}_{\bar{A}}$.

The Von Neumann entropy is frequently applied to probe nonequilibrium many-body physics. If we consider a one dimensional spin chain, the time-dependent growth of the half-chain entropy gives rise to the build up of bipartite entanglement in the system. While sufficiently local Hamiltonians, i.e., large values of the exponent $\alpha$, are associated with a linear entropy growth, this growth is only logarithmic for very long-range interactions \cite{lr_spread_schachenmayer}. The quasi-particle contribution, that is responsible for the linear growth \cite{qp_calabrese}, is increasingly suppressed by collective excitations with decreasing $\alpha$, leading to a slowdown of entanglement production \cite{lr_entgrowth_lerose}. Later on, we will see that the particular structure of quantum information is widely different in these scenarios.     

For many occasions, though, it is of particular interest how two subsystems $A$ and $B$ are correlated to each other, where $A\cup B\subset\Omega$ and $A\cap B=\emptyset$. For instance, two distant spins that are part of a larger spin chain. Based on the Von Neumann Entropy, one can define the \textit{mututal information}
\begin{align}
\label{mi}
    I_{A:B}=S_A+S_B-S_{AB}\,,
\end{align}
where $AB:=A\cup B$. One might be tempted to think that equation \eqref{mi} is a measure of entanglement between $A$ and $B$, however, the mutual information contains both, classical and quantum correlations, and is considered as a measure of total correlation between the two subsystems \cite{groisman_tot_corr}, see Fig\,\ref{fig:corr_measures}(b). The mutual information is strictly nonnegative, $I_{A:B}\geq 0$, where the equality holds iff $A$ and $B$ are uncorrelated, $\rho_{AB}=\rho_{A}\otimes\rho_{B}$. Furthermore, it is nonincreasing under reduction, i.e., $I_{A:BC}\geq I_{A:B}$. Given, for example, a many-body system out of equilibrium with interactions $\sim 1/r^\alpha$, Eq.\,\eqref{mi} can probe the aforementioned different regimes of information propagation associated with the exponent $\alpha$.

Even though the Von Neumann entropy \eqref{entropy}, or the mutual information \eqref{mi} offer vast insights into a quantum many-body system, they both quantify some type of correlation between two parties, i.e, \textit{bipartite} correlations. However, correlations can emerge among various parties, and it is far from trivial to grasp and quantify this \textit{multipartite} correlations. In particular, multipartite entanglement has evolved to be a field of research on its own over the last decades \cite{ent_horo}.

To probe correlations beyond the bipartite regime, we consider the \textit{tripartite mutual information} \cite{hosur_chaos} (TMI)
\begin{align}
\label{tmi}
    I_{A:B:C}=I_{A:B}+I_{A:C}-I_{A:BC}\,,
\end{align}
where $A,B,$ and $C$ are three disjoint subsystems of $\Omega$. The TMI was introduced in \cite{tmi_kitaev} as \textit{topological entanglement entropy}, where the authors used it to characterize multipartite entanglement in groundstates of topologically ordered two-dimensional systems. In the case of an overall pure state, the TMI \eqref{tmi} is symmetric under permutations of $A,B,C,$ and $D$, where $D$ refers to the complement of $ABC$. Moreover, $I_{A:B:C}=0$, if the systems state is separable with respect to \textit{any} partitioning of these subsystems, for instance: $\Ket{\Psi}=\Ket{\Psi}_{AB}\otimes\Ket{\Psi}_{CD}$. Thus, if Eq.\,\eqref{tmi} acquires a finite value, the systems' state is at least fourpartite entangled.   

From an information-theoretic point of view, the TMI \eqref{tmi} quantifies how quantum information distributes among the subsystems $A$, $B$, and $C$. Unlike mutual information, the TMI has no definite sign. While $I_{A:B:C}>0$ indicates that \textit{more} quantum information is shared among \textit{individual} subsystems, information is rather shared \textit{globally} if $I_{A:B:C}<0$. Furthermore, if the TMI has a large magnitude and a positive sign, \textit{local} measurements on $B$ and $C$ are sufficient to extract \textit{most} of the information shared with $A$. In the case of a negative sign, however, joint \textit{global} measurements on $BC$ are required. In the latter case, quantum information is said to be \textit{delocalized} with respect to the three subsystems, see Fig\,\ref{fig:corr_measures}(c). 

Moreover note that, if $I_{A:B:C}\leq 0$, mutual information is \textit{monogamous}, i.e.,
\begin{align}
\label{mono_mi}
    I_{A:BC} \geq I_{A:B}+I_{A:C}\,.
\end{align} 
Monogamy, a common notion in quantum information theory, is known to apply to quantum entanglement \cite{ent_mono}. Therefore, entanglement is not a shareable resource, implying that strong entanglement between $A$ and $B$ limits the amount of entanglement between $A$ and $C$. This is formalized by inequalities of the form $E_{A:BC}\geq E_{A:B}+E_{A:C}$, where $E$ is some measure of entanglement. Since mutual information is a measure of total correlation, it is not monogamous in general. However, delocalization of quantum information among $A$, $B$, and $C$ also implies the validity of the monogamy condition \eqref{mono_mi}. If the latter holds for arbitrary partitionings, one might argue that entanglement is the dominant correlation in the system \cite{corr_mono,tmi_holo_hayden}.

Recently, the TMI is applied to probe nonequilibrium physics, either in the context of unitary quantum channels \cite{hosur_chaos, tmi_schnaack}, or quantum many-body dynamics \cite{tmi_Iyoda,tmi_pappalardi,tmi_seshadri}. In the next section, we will examine how the exponent of the powerlaw decay affects the distribution of quantum information in the nonequilibrium dynamics of a quantum spin chain. Especially, we identify regimes of the exponent $\alpha$ that are associated with a sign change of the TMI at early times and, therefore, a qualitative change in the quantum information structure.

\section{\label{sec:deloc}Delocalization of quantum information}
To study how quantum information distributes over the degrees of freedom of a many-body system, we consider a one-dimensional chain of $N$ pairwise interacting spins (qubits) with open boundary conditions, described by the $XY$-Hamiltonian  
\begin{align}
\label{ham_xy}
    \mathcal{H}=\sum_{m<n}J_{mn}\left(\mathcal{X}_m\mathcal{X}_n+\mathcal{Y}_m\mathcal{Y}_n\right)\,\mathrm{.}
\end{align}
Here, $\mathcal{X}_m$ and $\mathcal{Y}_m$ denote the standard Pauli $X$- and $Y$-operators, acting on the lattice site $m$. We further choose the eigenbasis of the Pauli $Z$-operator as the local basis for each spin, $\mathcal{Z}_m\ket{0_m}=-\ket{0_m}$, and $\mathcal{Z}_m\ket{1_m}=\ket{1_m}$. The interaction strength $J_{mn}$ between two spins is given by a powerlaw
\begin{align}
    \label{powerlaw}
    J_{mn}=\frac{J_0}{\left\vert m-n\right\vert^\alpha}\,,
\end{align}
where $J_0$ is the nearest-neighbour interaction strength, and the exponent $\alpha\geq 0$ determines its spatial decay. Since the Hamiltonian \eqref{ham_xy} conserves the number of excitations, i.e. $\left[\mathcal{H},\sum_m\mathcal{Z}_m\right]=0$, the Hilbert space decomposes into a direct sum of invariant subspaces $\mathfrak{H}=\bigoplus_k\mathfrak{S}_k$, where each subspace $\mathfrak{S}_k$ is associated with a particular number $k$ of excitations. 

Eq.\,\eqref{powerlaw} results in nearest-neighbour interactions in the limit $\alpha\rightarrow\infty$. In this case, the $XY$ model \eqref{ham_xy} can be mapped onto a model of free fermions via a Jordan-Wigner transformation, and subsequent diagonalization in quasi-momentum space, i.e. 
\begin{align}
    \label{ham_nn}
    \mathcal{H}=\sum_k\epsilon^{}_k\eta_k^\dagger\eta_k^{}\,,
\end{align}
where $\eta_k^{\left(\dagger\right)}$ annihilates (creates) a fermionic quasi-particle with quasi-momentum $k$. The the Lieb-Robinson velocity is then determined by the maximal group velocity $v_\mathrm{LR}=\max\left(\frac{\mathrm{d}\epsilon_k}{\mathrm{d}k}\right)$, where $v_\mathrm{LR}=4J_0$ for the model at hand.

In what follows, we explore the delocalization of quantum information, measured by the TMI \eqref{tmi}, in the nonequilibrium dynamics induced by the Hamiltonian \eqref{ham_xy}. Initially, the system is in a product state $\Ket{\Psi_0}$ that is not an eigenstate of its Hamiltonian. We are, hence, interested in the nonequilibrium dynamics of the many-body state
\begin{align}
    \label{psi_t}
    \Ket{\Psi_t}=e^{-i\mathcal{H}t}\Ket{\Psi_0}\,,
\end{align}
where $\hbar=1$. This scenario is well suited to be performed on current experimental platforms. In the following section, we will describe in more detail how one can test our findings in the laboratory.

We first consider a simple scenario with an initial state that has just one localized excitation, for instance, at lattice site $i$: $\Ket{\Psi_0}=\Ket{0\ldots 1_i\ldots 0}$. The symmetry of the Hamiltonian then confines the dynamics of the system to the smallest (nontrivial) subspace $\mathfrak{S}_1$, which has dimension $N$. Accordingly, the state of the system \eqref{psi_t} can at any time be written as a superposition of basis states of this subspace
\begin{align}
    \label{psi_t_1}
    \Ket{\Psi_t}=\sum_m c_m\Ket{m}\,\text{,}
\end{align}
where the $c_m$ are time-dependent coefficients, $\Ket{m}:=\Ket{0\ldots 1_m\ldots 0}$, and $c_m=\delta_{mi}$ at $t=0$. During the nonequilibrium dynamics, the initially localized excitation coherently disperses and correlations between different regions of the system emerge. 

To study the TMI in this scenario, we shall compute the entropy $S_A$ of some subsystem $A$ that consists of a set of lattice sites. If we take the state \eqref{psi_t_1} and trace out all sites that are not associated with $A$, we obtain
\begin{align}
    \label{1_ex_rho_a}
    \rho_A&=\sum_{m,n\in A}c_m^{}c_{n}^*\Ket{m_A}\Bra{n_A}+\sum_{m\notin A}\left\vert c_m\right\vert^2\Ket{\mathbf{0}_A}\Bra{\mathbf{0}_A}\,,
    \end{align}
where $\Ket{m_A}$ is the state of $A$ with excitation at site $m$ and $\Ket{\mathbf{0}_A}:=\bigotimes_{m\in A}\Ket{0_m}$ is the state with all sites that belong to $A$ in the zero state. If we define $p_A:=\sum_{m\in A}\left\vert c_m\right\vert^2$ and
\begin{align*}
 \Ket{\mathbf{1}_A}:=\left(p_A\right)^{-1/2}\sum_{m\in A}c_m\Ket{m_A}\,,   
\end{align*}
we can diagonalize Eq.\,\eqref{1_ex_rho_a} in accordance with
\begin{align*}
    \rho_A=p_A\Ket{\mathbf{1}_A}\Bra{\mathbf{1}_A}+\left(1-p_A\right)\Ket{\mathbf{0}_A}\Bra{\mathbf{0}_A}\,\text{.}
\end{align*} 
The entropy of $A$ is then given by the \textit{binary entropy} 
\begin{align}
    \label{bin_ent}
    H\left(p_A\right)=\left(p_A-1\right)\log\left(1-p_A\right)-p_A\log\left(p_A\right)\leq 1\,,
\end{align}
and is, thus, tighter bounded than Eq.\,\eqref{vne_bound_gen} if $A$ contains more then one lattice site. The amount of entanglement between $A$ and its environment is, therefore, restricted to one 'e-bit' independent of the size of $A$.

Moreover, given two disjoint subsystems $A$ and $B$, it follows straightforwardly that the entropy of the union $S_{AB}$ is determined by $H\left(p_A+p_B\right)$. The TMI \eqref{tmi} then takes the form 
\begin{align}
    \label{tmi_bin}
    &H\left(p_A\right)+H\left(p_B\right)+H\left(p_C\right)+H\left(p_A+p_B+p_C\right)\nonumber\\
    -&H\left(p_A+p_B\right)-H\left(p_A+p_C\right)-H\left(p_B+p_C\right)\,.
\end{align}
According to Eq.\,\eqref{tmi_bin}, the TMI is a function of the variables $p_A$, $p_B$ and $p_C$, where $p_A$ can be interpreted as the probability of finding the excitation if we perform a measurement on $A$. At the boundaries of the parameter space, that is $p_A\lor p_B\lor p_C=0$, and $p_A+p_B+p_C=1$, Eq.\,\eqref{tmi_bin} vanishes, implying $I_{A:B:C}=0$. Furthermore, one can find a maximum of Eq.\,\eqref{tmi_bin} at $p_A=p_B=p_C=1/4$. Due to the concavity of entropy, we can then conclude that $I_{A:B:C}\geq 0$ for \textit{all} possible partitionings $A,B,C$. 

Thus, in this particular scenario, quantum information does not delocalize among spatial regions (in terms of the TMI), irrespective of the Hamiltonians' parameters. The effective size of the Hilbert space is just not sufficient for quantum information to spread properly over many degrees of freedom. This further implies that mutual information is either exactly extensive, i.e. $I_{A:BC}=I_{A:B}+I_{A:C}$, or nonmonogamous. Hence, entanglement can not dominate correlations overall, because the symmetry of the Hamiltonian strongly constrains the amount of entanglement that can emerge, see Eq.\,\eqref{bin_ent}. This result is in agreement with Ref.\,\cite{tmi_Iyoda}, where only for a few initial states with low effective dimension, a negative value for the TMI could not be observed. Noteworthy, delocalization of quantum information, and monogamy of mutual information require more than just nonseparability of the quantum state. For instance, it follows from the derivation above that generalized W states \cite{3qubit_duer}, which belong to one class of multipartite entanglement, will never lead to a negative TMI.

For a more sophisticated picture, we consider an initial state that leads to richer dynamics. This state is chosen to be a Néel ordered state $\Ket{\Psi_0}=\Ket{0101\ldots}$. In line with the groundstate of the model, this state is an element of the largest subspace $\mathfrak{
S}_{\left\lfloor N/2\right\rfloor}$. In that case, an analytical treatment is rather difficult, and we therefore resort to numerical calculations. In the spirit of the half-chain Von Neumann entropy, we divide the system into four connected regions $A,B,C,D$ of equal size, to probe delocalization of quantum information in general. In particular, we calculate the time evolution \eqref{psi_t} numerically, and evaluate the TMI \eqref{tmi} for this partitioning at every time step.

\begin{figure}
    \centering
    \includegraphics[width=.475\textwidth]{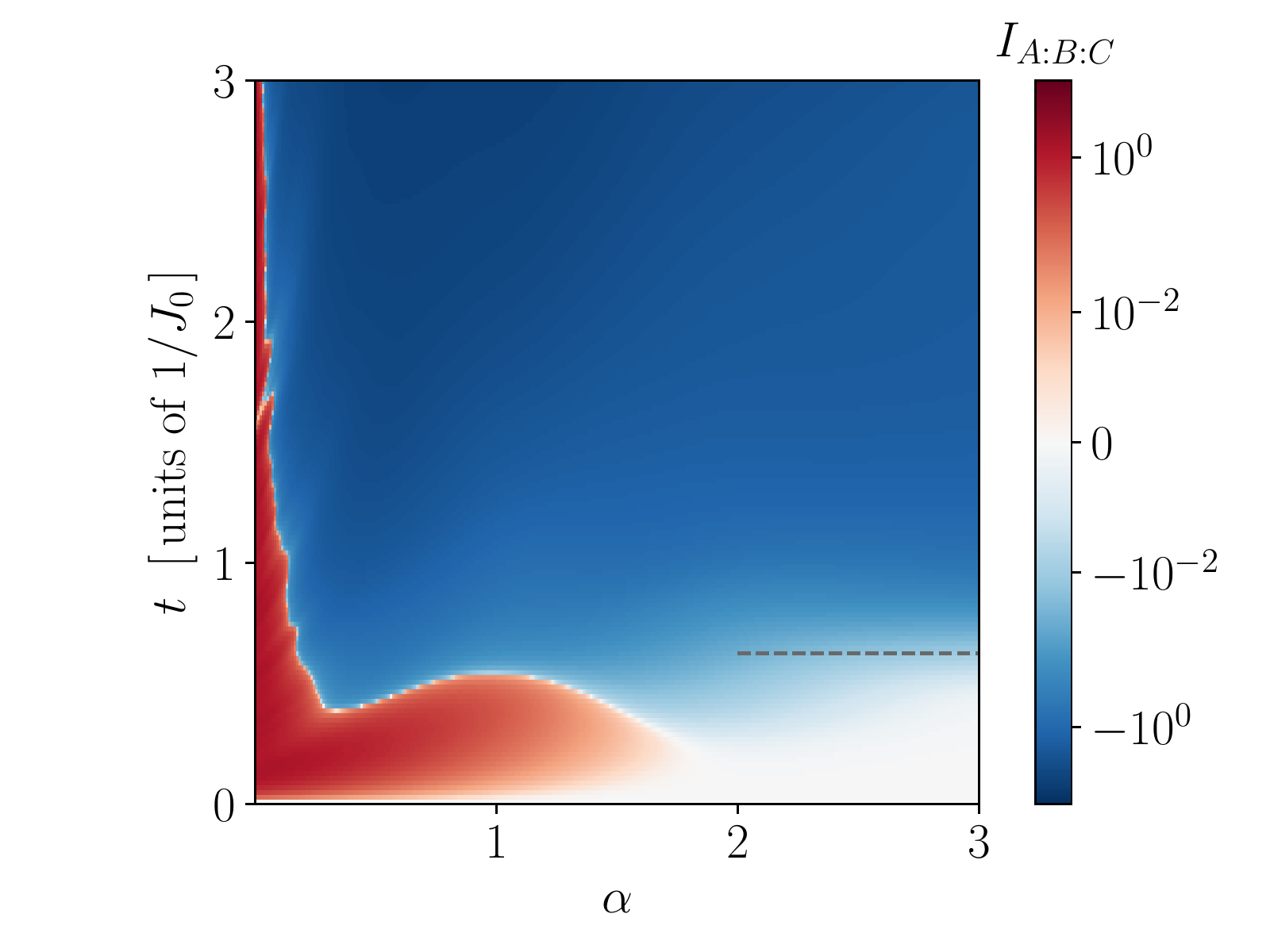}
    \caption{Nonequilibrium dynamics of the TMI \eqref{tmi} following from a Néel ordered state $\Ket{\Psi_0}=\Ket{0101\ldots}$ in the $\left(\alpha, t\right)$-plane for system of $N=20$ spins, and $A,B,C=\left\{1,2,3,4,5\right\},\left\{6,7,8,9,10\right\},\left\{11,12,13,14,15\right\}$. The dashed line shows the time it takes until the TMI can get sizeable in the nearest-neighbor limit, i.e. $\alpha\rightarrow\infty$, which is determined by the Lieb-Robinson velocity.}
    \label{fig:tmi_q}
\end{figure}

\begin{figure}
    \centering
    \includegraphics[width=.475\textwidth]{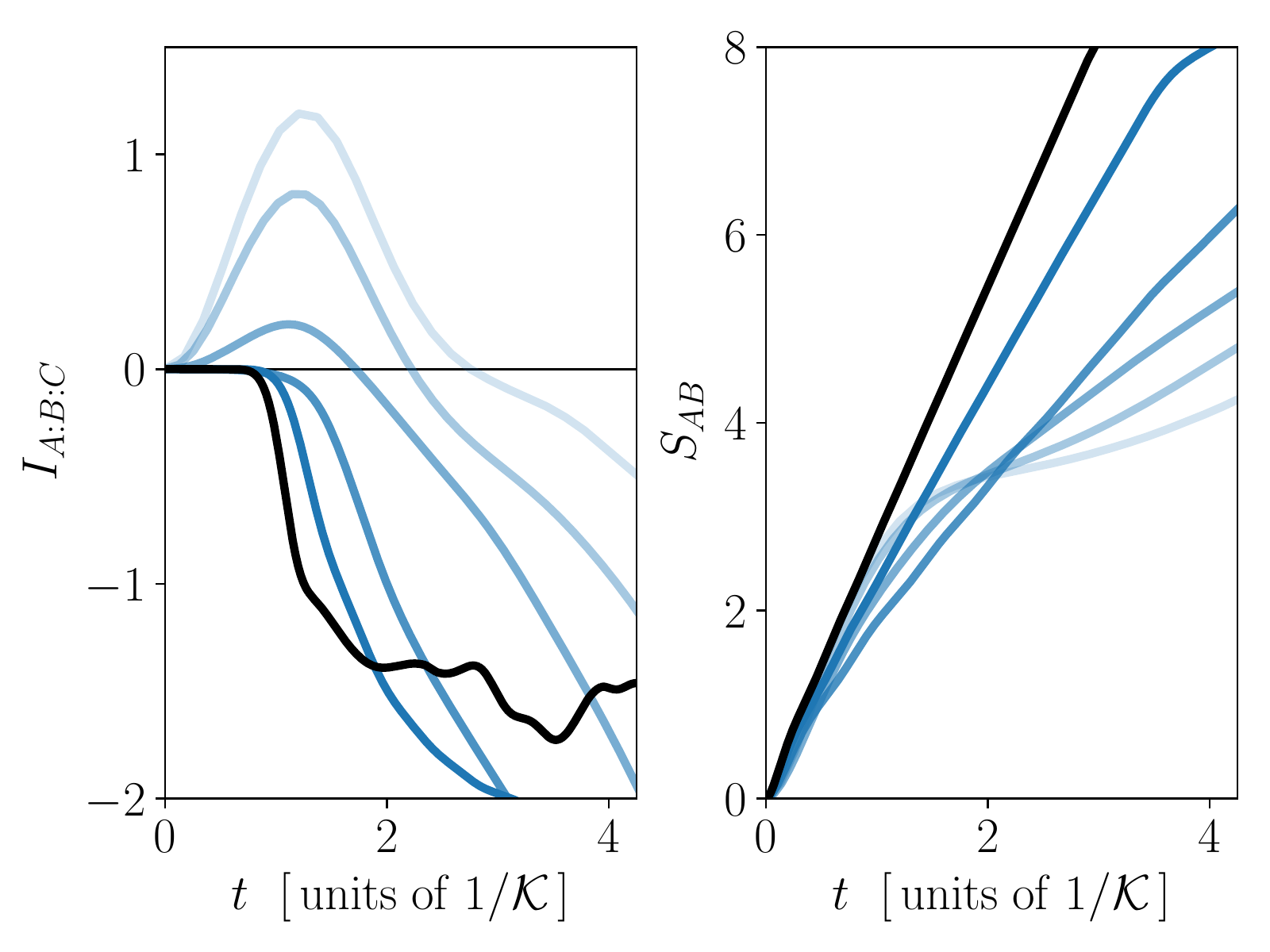}
    \caption{Nonequilibrium dynamics of the TMI \eqref{tmi} (left) and the half-chain entropy (right) following from a Néel ordered state $\Ket{\Psi_0}=\Ket{0101\ldots}$ for a system of $N=24$ spins, and $\alpha=0.3,0.5,1.0,2.0,3.0,\infty$ respectively. Time is rescaled by the Kac normalization $\mathcal{K}=\sum_{m<n}J_{mn}/N$. Darker color indicates larger values of $\alpha$. Each of $A,B,C,D$ is again a connected quarter of the chain.}
    \label{fig:tmi_q_s}
\end{figure}

In Fig.\,\ref{fig:tmi_q}, the TMI is displayed in the $\left(\alpha, t\right)$-plane for a system of $N=20$ spins. For larger exponents, i.e. $\alpha\gtrsim 2$, the TMI remains zero at early times and then attains a negative value. As this regime is associated with a clear lightcone \cite{lr_spread_hauke}, correlations between distant regions get exponentially suppressed at early times. The systems' state is, therefore, in good approximation still separable with respect to the subsystems $A$, $B$, $C$, leading to a vanishing TMI at early times. The time at which the TMI gets sizeable can be estimated via the Lieb-Robinson velocity from the nearest-neighbor version of the model (gray dashed line). The qualitative agreement of this estimate with the numerical data implies that the lightcone is (almost) linear in this regime. Once sizeable correlations among the subsystems start to build up, quantum information is delocalized, which follows from the negative value of the TMI. Oppositely, with increasing interaction range (decreasing $\alpha$), quantum information does not delocalize at early times. This effect, however, is only present at early times and the TMI also decays to a negative value for longer times, except for very small exponents, $\alpha\rightarrow 0$. Note that we observe similar behavior for other choices of the partitioning $A,B,C$. To put this result in perspective, Fig.\,\ref{fig:tmi_q_s} compares the TMI with the half-chain entropy for a system of $N=24$ spins, and particular values of the exponent $\alpha$. Here, time is rescaled by the Kac normalization \cite{kac} $\mathcal{K}=\sum_{m<n}J_{mn}/N$, to properly compare various values of the exponent $\alpha$. This rescaling fixes the average energy per spin independent of the exponent $\alpha$. On the shown timescale, the half-chain entropy is still in its growth phase. The linear growth in the nearest-neighbor case (black line) is increasingly suppressed to a logarithmic growth with decreasing $\alpha$. On the same timescale, the TMI undergoes a sign change for small exponents $\alpha$, i.e., an initial growth to a positive value following a decay to a negative one. The suppression of the positive peak at early times with increasing $\alpha$ can be nicely observed. Thus, for small exponents $\alpha$, the TMI clearly diagnoses a qualitative change in the quantum information structure at early times that is not covered by the half-chain entropy. These results demonstrate that the exponent of the powerlaw decay shapes the distribution of quantum information. Interestingly, regimes that are associated with a strongly altered causal region do not delocalize quantum information at early times.

\begin{figure}
    \centering
    \includegraphics[width=.475\textwidth]{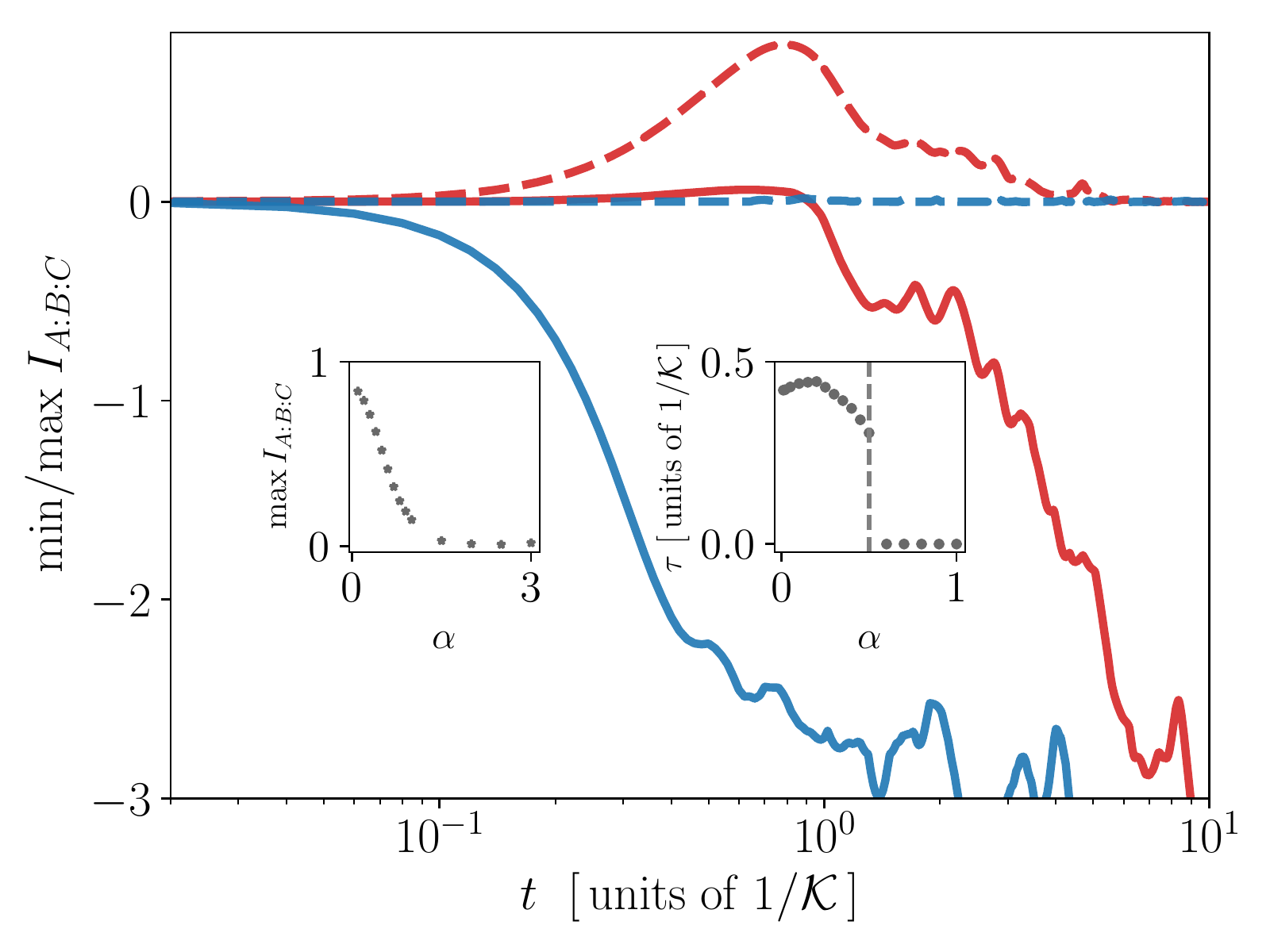}
    \caption{Maximal (dashed) and minimal (solid) TMI obtained from evaluation of Eq.\,\eqref{tmi} for all possible partitionings $A,B,C$ for a system of $N=12$ spins. Red color (dark gray) is associated with $\alpha=0.2$ and blue color (light gray) with $\alpha=3.0$. Time is rescaled by the Kac normalization $\mathcal{K}=\sum_{m<n}J_{mn}/N$. The left inset shows the largest value of the maximal TMI in the time interval of the main plot in dependence of $\alpha$. The rapid decay with increasing $\alpha$ suggests monogamy of mutual information in the nearest-neighbor limit. The right inset displays the time $\tau$ at which the minimal TMI out of all possible partitionings $A,B,C$ acquires a negative sign.}
    \label{fig:tmi_minmax}
\end{figure}

At last, we want to shed more light on the general structure of correlations in the identified regimes of the TMI. To this end, we calculate the minimal/maximal TMI out of all possible partitionings $A,B,C$ for a smaller system of $N=12$ spins. In Fig.\,\ref{fig:tmi_minmax}, the result of this calculation is shown for $\alpha=0.2$, and $\alpha=3.0$ respectively. We observe that for $\alpha=3.0$ the maximal TMI stays close to zero, while the minimal TMI monotonically decreases until saturation. In addition, the largest value of the maximal TMI in the considered time interval decays rapidly as a function of $\alpha$ as it can be seen in the left inset of Fig.\,\ref{fig:tmi_minmax}. This suggests that mutual information is monogamous overall in the nearest neighbour limit, that is, the monogamy condition \eqref{mono_mi} holds for \textit{all} possible partitionings. As previously mentioned, this indicates that entanglement is the dominant correlation in the system. On the contrary, for $\alpha=0.2$, both minimal, and maximal TMI are positive and growing at early times. Hence, there is not a single partitioning $A,B,C$ that fulfills the monogamy condition \eqref{mono_mi} until the minimal TMI changes its sign. We further investigate this behaviour in the right inset of Fig.\,\ref{fig:tmi_minmax}, where the time $\tau$ it takes until the minimal TMI acquires a negative sign is displayed in dependence of the exponent $\alpha$. Surprisingly, one can observe that this time is finite for $\alpha\leq 0.5$, and vanishes for $\alpha>0.5$. Thus, for $\alpha\leq 0.5$, correlations are nonmonogamous in general for $t<\tau$, which signifies that entanglement is not the dominant correlation at early times for very long-range interactions. Interestingly, even for $\alpha=0.2$ the maximal TMI decays towards zero at later times, and we observe this behavior for all $\alpha$ expect the fully connected case $\alpha=0$. The data therefore suggests that if $\alpha\neq0$, entanglement will dominate correlations at late times and quantum information is in general delocalized (in terms of the TMI). The longer the interaction range, the longer it takes until entanglement dominates. 

Our interpretation of these results is as follows. For large exponents, i.e. $\alpha\gtrsim 2$, the nearest-neighbor contribution \eqref{ham_nn} is the dominant part of the Hamiltonian \eqref{ham_xy}. The systems' dynamics is then understood by the propagation of quasi-particles, entangling different regions of the system as their propagate. Information that is initially localized in some region is spread by these quasi-particles which move at different velocities. Thus, this information will disperse, leading in general to delocalized information among subsystems, and entanglement being the dominant correlation. With increasing interaction range (decreasing $\alpha$), the Hamiltonian becomes more symmetric, and collective excitations more dominant. In the fully connected case, $\alpha=0$, where all excitations are of collective nature, the Hamiltonian is fully permutation symmetric. Similar to the $1$-excitation subspace, the additional permutation symmetry reduces the effective Hilbert space of the dynamics such that quantum information can not spread properly over many degrees of freedom. Or, in other words: collective excitations spread predominantly redundant information. Accordingly, we do not observe delocalization of quantum information for $\alpha=0$. Note that in Ref.\,\cite{tmi_seshadri}, it has been shown that the TMI of permutation symmetric states is typically positive. We can, thus, understand the qualitatively different dynamics of the TMI for small exponents as a remnant of this permutation symmetry. Due to the strong collective excitations in this regime, the state of the system accesses only a small portion of the Hilbert space at early times. Thus, for very long-range interactions, it takes more time for quantum information to delocalize, and entanglement to dominate.

\section{\label{sec:exp}Experimental realization}
As current experimental platforms may be able to observe our findings, we shall briefly touch the experimental realization, and possible obstacles regarding the chosen scenario. Our focus is on ion traps here, albeit other quantum simulation platforms are suited as well, especially in the nearest neighbour limit. For a more detailed description of these platforms, see the references we mention in the introduction of this work. 

In a linear ion trap, a string of atomic ions is confined via harmonic potentials, and two internal states of each ion serve as an effective spin-1/2 degree of freedom (qubit). The vibrational modes of the ions mediate an effective spin-spin interaction between these spins. This effective interaction can then be shaped via laser or microwave pulses to follow a powerlaw decay in accordance with Eq.\,\eqref{powerlaw}. We shall note that the interaction strength in reality deviates from the powerlaw shape as one moves towards the edges of the ion string. However, it has been shown lately that interactions can be realized more appropriately by applying additional optical tweezers \cite{espinoza}. 

The Hamiltonian, and the dynamics following the initial states we consider here have already been realized in systems of trapped ions \cite{xy_spread_richerme,xy_spread_jurcevic,xy_20q_friis}. To probe the TMI in an experimental environment, it is necessary to determine the Von Neumann entropy of different subsystems. The straightforward approach to accomplish this is quantum state tomography \cite{tomo}, however, as this technique scales quite disadvantageous with system size, it is only applicable to very small systems. We are not aware of any technique that is able to determine the Von Neumann entropy, and in addition circumvents quantum state tomography. Recently, though, a technique to determine Renyi entropies was introduced \cite{zoller_measure_1,zoller_measure_2,zoller_measure_lab}, and numerical tests indicate that Renyi entropies will lead to qualitatively similar results. For cold atoms in optical lattices, there also exists a method to determine the second order Renyi entropy \cite{islam_measure_lab,kaufman_measure_lab}. However, this method requires two identical copies of the system.

Finally, we shall note that due to coupling to the environment, the state of the whole system is in general mixed  in a real experiment in contrast to our idealistic assumption of a pure state. Hence, the whole system has a finite entropy unlike the entropy of a pure state. Whether or not the purity of current trapped ion systems is sufficient to observe our findings is an open question, that might encourage to further test the limits of these platforms.

\section{\label{sec:diss}Discussion}

Our results demonstrate that, besides the speed of information propagation, the exponent of the powerlaw decay starkly influences the distribution of quantum information. In regimes with an almost linear causal region, i.e. $\alpha\gtrsim2$, we find that quantum information is mainly delocalized. Furthermore, the data shows that nonmonogamous correlations get strongly suppressed with increasing $\alpha$, which suggests that correlations are predominantly caused by entanglement in this regime. Intuitively, this is understood via the picture of fermionic quasi-particles that delocalize quantum information as they propagate. 

With increasing interaction range, the systems' Hamiltonian becomes more symmetric, and collective excitations more dominant. Hence, the state of the system accesses only small portions of the Hilbert at early times, protecting quantum information from delocalization. For sufficiently slow decaying interactions, $\alpha\leq0.5$, we find that mutual information is nonmonogamous overall for a finite time $\tau$. Note that for an intial product state, this regime is associated with distant independent propagation of information \cite{eisert_break_locality}. Thus, it seems that a strongly altered causal region is accompanied with sizeable nonmonogamous correlations at early times. Entanglement, therefore, needs some finite time to dominate the systems' dynamics in this case, which might be linked to semiclassical descriptions of long-range interacting systems \cite{lr_entgrowth_lerose,lerose_semicl}.

Despite these exciting results, many questions remain open. Although our results suggests that the structure of quantum information vastly differs in the identified regimes, so far we do not know how beneficial or disadvantageous these structures are with respect to quantum information processing. Moreover, as our calculations were carried out for relatively small system sizes, and a specific model, it is natural to ask how general these findings are. It would be fairly interesting if one could lay out analytical arguments that mutual information is monogamous in systems described by local Hamiltonians, as it is the case for quantum field theories with holographic duals \cite{tmi_holo_hayden}. At last, we want to emphasize that the TMI indeed probes  a many-body system beyond the insights given by the Von Neumann entropy, or mutual information. It would be compelling to see what insights this measure can provide regarding nonequilibrium phenomena like thermalization, and many-body localization \cite{rev_neq_eisert}, for example. We hope this work stimulates further efforts in this direction both theoretically, and experimentally.

\section*{Acknowledgements}
D.W. gratefully acknowledges support support from the Helmholtz Institute
Jena and the Research School of Advanced Photon Science of
Germany.

\bibliography{ref}
\end{document}